\documentclass[smallcondensed]{svjour3}

\smartqed  

\usepackage{graphicx}
\usepackage{mathptmx} 
\usepackage{hyperref}
\usepackage{color}  

\journalname{Journal of Grid Computing}

\begin{document}

\title{Building the International Lattice Data Grid
}

\titlerunning{International Lattice Data Grid}

\author{Mark~G.~Beckett \and
        B\'alint~Jo\'o \and
        Chris~M.~Maynard \and
        Dirk~Pleiter \and
        Osamu~Tatebe \and
        Tomoteru~Yoshie
}


\institute{M.~Beckett and C.~Maynard \at
 The University of Edinburgh, Edinburgh, United Kingdom \\
 Tel.: +44-131-650-5030\\
 Fax: +44-131-650-6555\\
 \email{george.beckett@ed.ac.uk c.maynard@ed.ac.uk}
\and B.~Joo \at
Scientific Computing Group, Jefferson Lab, \\12000 Jefferson Avenue, Newport News, \\VA 23606, U.S.A
\email{bjoo@jlab.org} 
\and D. Pleiter \at
  Deutsches Elektronen-Synchrotron DESY, 15738 Zeuthen, Germany \\
  Tel.: +49-33762-77-381\\
  Fax: +49-33762-77-216\\
  \email{dirk.pleiter@desy.de}
\and T.~Yoshie \at
  Center for Computational Sciences, University of Tsukuba,
  Tsukuba 305-8577, Japan \\
  Tel.: +81-29-853-6492\\
  Fax:  +81-29-853-6406\\
  \email{yoshie@ccs.tsukuba.ac.jp}
}

\date{Received: date / Accepted: date}

\maketitle

\begin{abstract}
  We present the International Lattice Data Grid (ILDG), a loosely
  federated grid of grids for sharing data from Lattice Quantum
  Chromodynamics (LQCD) simulations. The ILDG comprises of metadata,
  file format and web-service standards, which can be used to wrap
  regional data-grid interfaces, allowing seamless access to catalogues
  and data in a diverse set of collaborating regional grids. We discuss
  the technological underpinnings of the ILDG, primarily the metadata
  and the middleware, and offer a critique of its various aspects with
  the hindsight of the design work and the first full year of
  production.

\end{abstract}

\section{Introduction}
\label{intro}
In this paper, we present the International Lattice Data Grid (ILDG). The ILDG project is a mostly volunteer effort within the Lattice Quantum Chromodynamics (LQCD) community, to share data worldwide, and to thus amortise the very high computational cost of producing the data. In terms of organisation it is a data-grid, but it is also a loosely federated grid of grids. 

Large data sets require significant scientific endeavour to amass
them.  This may represent intellectual property, as well as physical
resources.  In the case of LQCD, the resources are both intellectual --
such as the scientific ideas and algorithmic development -- as well as
other resources, such as the manpower required to write the computer
code and the resources to procure/develop and operate a large
supercomputer. Why then do scientists wish to share this valuable
data? It is precisely because this data is so valuable that scientists
make it available for others to use. A mechanism is required whereby
those who generate shared data can receive credit for doing so.

For the LQCD community there are two compelling reasons to
share data.  First, fully exploiting the data requires computing and
manpower resource. A particular group may generate a dataset to
compute a target physical quantity with sufficient precision to have
an impact on experimental results, and yet not have sufficient
resources or even the expertise to calculate many other possible
quantities on that dataset. At this stage, rather than waste some of
the scientific potential of the data, a group may give the data away
freely provided some basic use conditions are met such as citing a
certain paper in any resulting publication. Second, the resources
required to generate ever more potent data sets require ever greater
resources, outstripping Moore's Law and scientific innovation. This
forces different groups to collaborate: jointly baring the cost of
data generation.

Quantum Chromodynamics (QCD) is a theory of sub-atomic particles (specifically, quarks and gluons) and their interactions. Lattice QCD (LQCD) is a version of QCD where space-time is discretized, making the theory amenable to calculation by computers. LQCD computations are of utility in a variety of theoretical particle physics contexts including Nuclear Physics and High Energy Particle Physics, and have historically consumed a large fraction of available computing cycles worldwide. The interested reader can find several excellent books and review articles on LQCD in the literature, for example \cite{Creutz:1984mg,Montvay:1994cy} and \cite{Gupta:1997nd}.

LQCD Computations are based on Markov Chain Monte Carlo methods (see
\cite{Peardon:2003fv} for a recent review) and typically the primary
data from such calculations are samples of the QCD vacuum known as
{\em gauge configurations.} The Monte Carlo process will generate an
{\em ensemble} of such configurations for each set of physical and
algorithmic input parameters. At the time of writing, the typical cost of
generating an ensemble is $O(1)-O(10)$ Teraflop years depending on the
precise formulation employed, and this cost is expected to grow to the
Exaflop-year scale as one simulates lattices with finer lattice
spacings, larger physical volumes and physically light quarks.

A set of ensembles is amenable to many different type of secondary
analysis. One can, for example, perform calculations of nuclear
structure on the same configurations one also uses to perform calculations
of fundamental parameters of the Standard Model of Particle
Interactions. Alternatively, an ensemble generated to measure Nuclear
Energy spectra may also be useful in the study of the nuclear strong
force binding together nucleons into atomic nuclei.

Since the generation of ensembles is very demanding in terms of
effort, and since the ensembles can facilitate multiple uses, it makes
sense to share them amongst the LQCD community to get maximum value
out of a particular generation project. The ILDG infrastructure 
discussed in this paper, is designed to promote and facilitate such data sharing.

The LQCD community has a history of sharing data before the formation of the ILDG.  The MILC collaboration~\cite{MILC} has pioneered the approach of freely giving away the data, after publishing results for their target quantities. This conservative approach is necessary for scientific prudence. The data has been very widely used, and the MILC collaboration policy of data release is seen as successful and beneficial to the collaboration.  There are many examples of different groups collaborating together to share the burden of generating the data. In some sense the ILDG is similar to other kinds of data archives and Science Gateways, of which there are now many throughout the world. However, it does present some particularly unique aspects, which stem from it being a {\em grid of grids}.  

In 2002, different groups were starting to make use of grid
technologies to store and retrieve data, primarily within their own
collaboration. A proposal by Richard Kenway~\cite{Davies:2002mu},
at the annual lattice QCD conference, to use grid technologies to
store and share data, was well received and supported.  The ILDG was
formed from interested groups that were willing to share data. There is
no central authority forcing policy on the member collaborations:
rather the ILDG is a collaboration of groups that are prepared to
commit some resource to a central service. This idea of an
aggregation or grid-of-grids is a powerful one, which allows each
group to retain control of its own resources whilst making them
available to the greater whole.

This paper is organised as follows: we outline the basic
requirements needed for such an infrastructure in section
\ref{S:requirements}.  The ILDG development has been split into two
broad overlapping groups, the metadata working group (MWG) and the
middleware working group (MWWG).  In sections \ref{S:metadata} and
\ref{S:middleware}, we consider aspects of metadata and middleware,
respectively. Finally, in section \ref{S:review}, we review aspects
of the ILDG project from over several years of activity and over one year of
production. We present operational details of the
infrastructure as well as criticism of various aspects. We also
have a chance in section \ref{S:review} to compare and contrast the
ILDG with some related or similar efforts. Finally, we summarise and
discuss the potential for future work in section \ref{S:conclusions}.

\section{Requirements}
\label{S:requirements}

%

Put in simple terms, the goal of the ILDG project is to allow scientists
to share their data, across the different research collaborations within
the project.

In order to attain the goal, the team have had to translate it into a
set of concrete requirements, which has then been used to guide the
development of the ILDG infrastructure (that is, the technologies,
policies, and processes).  These requirements are summarised in this
section.

\subsection{Data management} 

The team started by quantifying the data to be shared. This data is
file-based and -- as noted above -- represents lattice gauge
configurations, which are collected together into {\em ensembles}
pertaining to Monte Carlo simulations.

The nature of the simulations implies that a configuration is only
meaningful as a member of the ensemble. Thus, scientists almost always
want to access the whole ensemble (or a significant part thereof): this
equates to Terabytes of data.  Also, scientists typically need to have
access to local copies of data, in order to complete the required
analysis processes. Thus, it is clear that {\em sharing data} involves
copying multi-Terabyte file sets from the storage facility of one
research group to a remote scientist's local system.

All file copy operations are intended to be undertaken over the
Internet. Thus, even with good bandwidth, it is clear that
multi-Terabyte transfers represent time-consuming operations, requiring
a reliable, high performance bulk data transfer mechanism.

Ensembles of gauge configurations that pre-date ILDG are typically
identified using locally agreed naming conventions. For example, a
particular configuration might be identified by a combination of the
Unix path to the file and the hostname of the server on which it
resides. While this approach may be suitable for a small group of
researchers working in a particular collaboration, it is inadequate for
a community like ILDG that is loosely coupled and distributed across
multiple research groups.

What is required is a method for assigning a {\em unique and persistent
identifier} to each file (that is, gauge configuration) that is to be
held within the infrastructure. In addition, there needs to be an
equivalent method for identifying each ensemble.

\subsection{Data Curation}

For a configuration (or an ensemble) to be useful to a researcher, it
must be apparent what it represents in scientific terms. This
information is provided by {\em metadata} -- literally, data about
data.  Metadata may be captured in a number of different ways. For
example, a widely used approach is based on descriptive filenames that
follow an agreed naming format.  For Lattice QCD, the detail required
to describe a dataset is too great to be realistically encoded in its
filename, especially considering the various different formulations of
QCD available, all with different parameters. The process of scientific
annotation has warranted a more sophisticated approach.

ILDG researchers require a scientific annotation that thoroughly and
unambiguously describes a configuration (or ensemble of) for other
members of the community. The annotation should be extensible: that is,
it should support the introduction of new descriptive elements. This may
be required --- for example --- to accommodate new science.

A user should easily be able to search the catalogue of scientific
annotations and, complementing this, the generation of metadata should
be a lightweight and straightforward process. Where possible, elements
of the description should be populated automatically.

As well as having an agreed mechanism for describing data, one must also
be able to read the binary files that hold the data. This motivates
convergence to a common file format (for gauge configurations, at
least).  At the inception of ILDG, a number of different file formats
existed, based on the conventions used in the most popular LQCD
codes. Alongside the formalisation of the scientific metadata, it has
been decided that a community-wide, flexible, extensible binary file
format is required.

\subsection{Infrastructure}

Pre-dating the formation of ILDG, the five collaborations that make up
the core of the consortium have procured or developed storage facilities
to host the ensembles of data that they each generate. These systems are
all accessible, in principle, over the Internet, but via different and
incompatible access protocols and access control systems targeted at
local (that is, institution-based) users. 

To work around this issue, two specific requirements need to be
fulfilled. First, a layer of software is required on top of the local
infrastructures, to provide a uniform interface to an end-user. Second,
an access control mechanism needs to be established that permits ILDG
members from different collaborations to access designated data at
partner institutions/storage facilities.

\subsection{Operation and monitoring}


To be useful, the ILDG infrastructure must achieve high levels of
availability. High availability must be attained in spite of the
decentralised and heterogeneous nature of the component elements, and
should efficiently exploit the support effort available at the regional
grids. It has therefore been decided that an automated monitoring
service should be set up within the infrastructure, fulfilling the
following specific attributes. The monitoring service needs to:

\begin{itemize}

\item be reliable -- since it is the primary means in which problems and
  failures are identified.

\item be flexible -- in order that the diversity of ILDG components can
  be represented and monitored.

\item produce accurate and informative alarms, which will allow
  regional-grid support teams to quickly and effectively diagnose and
  resolve issues.

\item post alarms using email -- as this is the primary medium over
  which regional-grid teams communicate.

\item maintain a record of system performance, to inform coordinators as
  to overall reliability and to highlight any specific weaknesses.

\end{itemize}

For easy access to all ILDG resources, a centrally coordinated user
management system is required: making all globally registered users
known to all local-resource providers. To this end, we have adopted the
concept of a Virtual Organisation (VO), with membership being managed by
the VO itself. With ILDG consisting of several regional grids, a setup
is however needed that allows the decision -- as to whether an
application for VO membership is to be approved or declined -- to be
delegated to the regional grids. For users to access ILDG resources only
a single sign-on should be required: that is, a single trust domain has
to be defined. This domain should include a sufficiently large set of
trusted Certificate Authorities that every potential users can be
provided with a certificate that is acceptable to any of the resource
providers.

While it is not envisaged that the regional-grid make-up of ILDG will
change in a particularly dynamic manner, it is expected that new
collaborations will wish to join the infrastructure, either
independently or as part of an existing group. With this in mind, it is
important that the infrastructure evolves in a way that does not prevent
expansion. Specifically:

\begin{itemize}

\item ILDG specifications (for example, service definitions) are
  thoroughly documented in a manner intended to facilitate the creation
  of new implementations.

\item the technology layer is supported by a test suite, which allows
  new implementations of ILDG services to be validated against the specification.

\item where possible, ILDG uses open (or at least widely adopted)
  technologies and standards, aiming to increase coverage of user groups
  and to reduce the risk of systems becoming obsoleted.

\item the technological aspect of the infrastructure is specified as a
  thin layer (that is, focused on a baseline set of functionalities),
  which can easily be incorporated into existing infrastructures with
  low levels of development effort.

\end{itemize}

\section{Metadata}
\label{S:metadata}

To motivate the need for metadata, consider an example where there is no
metadata. Configurations from different ensembles are all stored in a single
directory with potentially random strings for names. Clearly this data is now not accessible.
A scheme is required to describe the data. As noted above, many groups have in the past constructed 
ad-hoc schemes for describing the data based on filenames and directory structures. Whilst this approach is not without merit, it does not scale when many 
groups are sharing data. In constructing this scheme there are likely to be several assumptions 
which are specific to the group which uses the scheme. Another group 
may well find these assumptions are not valid for their data, and hence their
data will not fit into the scheme. Modifying the scheme is only possible where 
the assumptions used in its construction are still valid. To accommodate several
potential different formulations of LQCD, and the needs of different groups
a different approach is required.

Extensibility is a critical requirement of any annotation scheme. Any new data will need 
new metadata to describe it and the scheme will have to be modified. In an extensible scheme this
can be done without breaking the original scheme. That is, the new scheme is an 
extension of the old one. Furthermore, any document which was valid in the old scheme is 
valid in the new one, so that the old documents don't have to be updated to be valid in the new scheme. 

Data provenance is likewise an important requirement. Can the data be recreated from the metadata? 
Taken to the limit this question is extremely challenging. In principle the code used to generate the data and its inputs should allow 
the data to be regenerated. However, this doesn't include any machine, compiler or
library information. Moreover, in the context of sharing data, the application belonging to one group may not be able to 
parse and process the input parameters of the application belonging to a different group. Hence while 
a full archival of a statically linked code,  its inputs should allow recreation of the
data if the original producing machine were to be available, archiving to this level of detail
is not practical.  Correspondingly some of the data provenance requirements may need to be 
softened in practise.

Lattice QCD metadata is hierarchical in nature and the annotation
scheme should reflect this.  Markup languages combine text and
information about the text, and thus are perhaps a natural choice for
a language in which to construct the scheme. Semantic or descriptive
languages don't mandate presentational or any other interpretation of
the markup. XML was chosen as it is the most widely used and best
supported markup language. Similarly XML schema was chosen as the
schema language to define the scheme or set of rules for the metadata.

In order to make sharing lattice QCD data useful and effective,
lattice QCD metadata should be recorded uniformly throughout 
the grid. The metadata working group designed an XML schema called
QCDml for the metadata. The primary use case being data discovery via
the metadata. 

As described above a key concept for Lattice QCD data is the organisation of the 
data as configurations and ensembles to which the configurations belong. The metadata
is divided into two linked XML schemata, one for the configurations, and one for the ensemble. 
The two schemata are linked together by a unique Uniform Resource Identifier (URI), called the {\tt markovChainURI}, which lives in 
the name-space of the ILDG and which appears in the XML instance documents (IDs) of the configuration 
and the ensemble to which it belongs. There is no formal mechanism for ensuring uniqueness, 
but a simple convention has been adopted whereby the name of the group who generated the data 
appears in the URI, and responsibility for uniqueness is thereafter delegated to that group.

The separation of the metadata into two pieces, besides reflecting the nature of lattice QCD data, 
has two advantages. First, metadata capture is potentially simplified, as only the configuration-specific information has to be recorded for each configuration, and the information specific to 
an ensemble has to be recorded only once. Second, the performance of searches on the data may be improved 
since the split represents a factoring of the original more complicated schemata.
 
The metadata scheme is encoded as a set of XML schemata
\cite{XMLSchemata} and whilst this does not mandate how the metadata
is stored and accessed, for simplicity it is often stored in native
XML databases such as {\em eXist} \cite{eXist}. It is well known that
the speed of access of hierarchical databases, such as native XML
database is vastly inferior to that of relational databases. 
Scientists are almost always interested in finding an ensemble rather
than finding an individual configuration. Therefore, for most cases,
the separation of ensemble and configuration XML reduces the number
of documents to be searched by ${\mathcal O}(100-1000)$.

In each configuration ID the {\em logical file name} (LFN) of the data
file is recorded. The LFN is a unique and persistent identifier of the file
in the ILDG name-space. The ILDG and local grid services then map the LFN to
actual file instances. 

The data itself is stored in a file format known as LIME~\cite{LIME}. LIME is short for Limited Internet Message Encapsulation, and is a simplified and generalised version of the DIME (Direct Internet Message Encapsulation) \cite{DIME} Internet standard, which was proposed as an Internet Standard and which is now part of the Microsoft .NET framework. LIME is a record-oriented message format which simplifies and extends the original DIME framework by introducing 64-bit length records instead of the original 32-bit ones, and correspondingly it eliminates the need for continuation records. LIME thus allows the packing of descriptive text records and binary data records in the same file. This format itself is very flexible and extensible since the types and sequence of records are not mandated in the file format itself. The ILDG however, specifies and requires a set of LIME records, including: a record containing some XML file format metadata describing the size of the space-time lattice and data precision; a record containing the data itself in a specified data ordering; and a record of the LFN for the data, to allow the linking of ILDG data files to their metadata catalogue entries. LIME was developed by the USQCD collaboration through the SciDAC software initiative and a C-code to read and write lime files on a serial machine (C-LIME) can be downloaded from the USQCD web-site \cite{LIME}. The QIO package also developed by the USQCD collaboration has facilities for reading ILDG formatted data files on both serial and parallel machines \cite{QIO}

The scientific core of the metadata is contained within the ensemble schema.  
The most important section from the data discovery viewpoint is the {\em action} 
which contains the details of the physics. Here, the object-oriented ideas of inheritance 
are used to build an inheritance tree of actions based on the XML schema concepts of 
extension and restriction as appropriate. This enables users to make both very specific 
searches and more general searches on the names, types
and/or parameters of the actions. The exact details of the physics can only
be encoded in mathematics, which is not suited to an XML description. 
A reference to a paper, and the URL of an external {\emph glossary} document
which contain the mathematical descriptions of the physics are included. Clearly
an application cannot parse this information, but it is included to avoid 
ambiguity in the names used in the inheritance tree.

%
%
%

QCDml uses a namespace defined by an URI. This URI includes the version number. Backward compatible, extensible updates to the schema don't change the URI of the namespace, so XML IDs don't need to be
modified. Clearly the XML ID of the schema itself is modified, so a new URL for the extended XML ID of the schema is needed. All versions persist on the web, but with incremental URLs. Non-extensible updates to the schema require a change of namespace and the URI which identifies it.

Lattice QCD algorithms are very complex with many different
algorithmic components.  They are also an active area of research, and
changes and improvements are common. This makes designing a scheme,
especially an extensible one rather difficult. The defined names and
inheritance tree ideas used for the action would be too cumbersome for
describing the algorithms. QCDml has only small scope for the
algorithms limited to name, value pairs for the
parameters. Algorithmic details can be expressed mathematically in an
external, non-parseble glossary document.  This approach further
limits the data provenance of QCDml. However, individual groups can
import their own namespace with as much detail and structure as they
see fit, which can help ameliorate the data provenance issue even if
the metadata is no longer universal.

Before leaving this general discussion of the metadata schema we note
that the current set of schemata may be found at \cite{EnsembleSchema} (ensembles) and \cite{ConfigurationSchema} (configurations). More Physics-oriented information 
about the metadata can be found in \cite{Coddington:2007gz}.
\section{Middleware}
\label{S:middleware}

As described above a grid-of-grids concept has been adopted for ILDG.
Each of the regional grids has to provide the following services: a
metadata catalogue (MDC) for metadata-based file discovery, a file
catalogue (FC) for data file location and one or more storage elements
(SE) which can then serve the data to the user.

The user can discover available datasets by sending a query to the
MDC of each of the regional grids. On input this search requires an XPath
expression. On output the services will return the list of Logical Files Names
(LFNs) of those documents for which the XPath expression identifies a
non-zero set of nodes. 

To identify all copies of a particular file a query to the file
catalogue has to be performed, which takes an LFN on input and returns
a source URL (SURL) for each replica of the file.  The scheme part of
the SURL tells the client whether it can either directly download the
file using the transfer protocols HTTP or GridFTP or whether it has to
connect to a Storage Resource Manager (SRM) interface. 

The SRM protocol \cite{srm} is evolving to an open standard for grid middleware to communicate with site-specific storage fabrics.  The ILDG, at the time of writing, requires the SRM to be adhere to version 2 of the SRM specification. A particularly appealing feature of the SRM is that implementations of the service are typically provided with a standard Web-Service interface, allowing the SRM component to fit in easily with the comparatively less complex, ILDG-defined MDC and FC.

SRM is a system to manage local storage fabrics comprised of several
data servers and possibly long-term backup storage. From the point of
view of an ILDG transaction, the SRM may be required to: stage a file
to some transfer location on a server, negotiate a transfer protocol
between the server and the client, and to then arrange for the
transfer to occur. Once the file is ready for transfer, the location
and the transfer protocol are returned to the client by the SRM service
in a transfer URL (TURL).  The client and the data server can then
carry out the transfer independently of the rest of the SRM
system. Typically GridFTP is used as a transport protocol.

By adopting a grid-of-grids concept with different middleware stacks
being used by the different regional grids, interoperability becomes
a challenge.  Interoperability is required to provide standardised
interfaces towards the application layer. While there are clear
similarities between the different grid architectures, there are crucial
conceptional differences and incompatibilities of the interfaces.

For all services, which have not been specifically designed for ILDG,
two strategies have been applied to overcome this interoperability
issue. Firstly, wherever possible, common grid standards supported by
all the used middleware stacks have been adopted. One example is the
transfer protocol GridFTP.  Secondly, interface services have been
defined and implemented. Instead of accessing a service directly, the
user connects to the interface service which will process the request
on behalf of the user.

If a service requires authentication the corresponding interface
service has to provide a credential delegation service. Within ILDG
we use an implementation of such a service that has been developed
within the GridSite project \cite{gridsite} and is now part of the gLite
middleware stack.  On request of the client the server returns a proxy
certificate request, which is signed on the client side and returned back
to the server.  Since the proxy certificate has only a limited lifetime,
the risk due to a compromised server hosting the interface service is
considered to be acceptable small.

To standardise a web-service within ILDG a WSDL description is
implemented and additionally a behavioural specification is provided.
The WSDL description specifies the structure of the service's input and
output data structure, while a functional description of the service is
provided by the behavioural specification. Additionally, test suites
have been defined and implemented which can be used to verify whether
a service conforms to the ILDG standard.

Access to most services is restricted to members of the Virtual Organisation
(VO) ILDG. For the management of this VO we use a VOMRS service \cite{vomrs}.
Each user which wants to join the VO has to submit an application and
nominate one of their regional grid's representatives. For each regional
grid at least two representatives have been assigned which can accept
or reject the request. For each regional grid an individual group has been 
created. Information on group membership may be used by the regional grids
as input for authorisation services.

The only other global service which is used within ILDG is the monitoring
service. This monitoring service has been implemented using INCA \cite{inca}. 
In the framework of INCA a set of so-called reporter managers regularly
execute test scripts accessing grid resources. The information returned by
the reporter managers is collected in a repository. In case of failures
a notification email is generated and sent to the regional grid which is
responsible for a particular service. Data in the repository can later
be used to check the service's availability.

%

\section{Review}
\label{S:review}

\subsection{General Status}
%
%
%
At the time of writing, the ILDG has been in production use for a little
over a year.  It is comprised of five main-partner regional grids. These
are: The Center for the Structure of Subnuclear Matter (CSSM) in
Australia; The Japan Lattice Data Grid (JLDG); Latfor Data Grid (LDG)
for continental Europe (primarily Germany, Italy and France); the
regional grid of the UKQCD Collaboration in the UK; and the regional
grid of the USQCD collaboration in the United States.  The ILDG VO has
113 registered users and the combined ILDG hosts some 207 gauge
configuration ensembles, corresponding to various lattice volumes, gauge
and fermion actions. Each single ensemble represents significant
portions -- potentially years -- of human and supercomputing
resources. Thus these archives are immensely valuable.

On the management side, the Middleware working group hosts monthly
teleconferences to discuss operational exceptions, experiences and
future development efforts while at the higher level, the ILDG holds
bi-annual video conferences, so that regional partners can discuss
more general progress.

\subsection{Benefits of Sharing}
%
%
%
Hosting such a wealth of data has had great benefit on computational
lattice QCD worldwide. In the case of some regional grids, the regional
grid itself has become the primary means of data distribution, for multi-site
projects, prime examples of which are the LDG and UKQCD
collaborations \cite{Beckett:2009}.

A number of research activities have been enabled, thanks to the ILDG infrastructure. Scientists in Japan have been using data produced by the MILC collaboration (in United States) as part of their research \cite{Furui:2008um} and, complementing this, a team at $\chi$-QCD (University of Kentucky) has accessed data from CP-PACS (Japan), in the ILDG community \cite{Draper:2008tp,Doi:2009sq}. Other examples of ILDG use can be found in \cite{Ehmann:2007hj} and \cite{Ilgenfritz:2006gp} where two groups made use of lattices generated by the German QCDSF Collaboration. 

Both inter-collaboration and intra-collaboration activities are enabled by ILDG. In \cite{Stuben:2005uf}, a number of ILDG-enabled activities are noted relating to data sharing across LDG sites. The fact that the ILDG is making a serious impact in international collaboration can also be seen in the fact that Physics workshops are being held within the community that focus, not only on the generation of QCD data, but also on accomplishing calculations by sharing the data via the ILDG \cite{TsukubaWorkshop}.

\subsection{Criticism of the ILDG}
While ILDG appears to be operating successfully, there are some aspects of it that could be improved. Using the ILDG to locate and share data is relatively straightforward especially with the easy to install client tools \cite{ILDGTools} as is described in \cite{Yoshie:2008aw}. However, contributing to the ILDG potentially involves a lot of effort. Depending on the level of involvement, one may need to maintain storage and database resources as well as having to mark up configuration and ensemble metadata.

In order to create ensemble metadata markup, one needs to get a unique key to identify it ({\tt MarkovChainURI}).  There is no service which can supply one or necessarily check that a manually chosen key is in fact unique. Further, ensemble metadata markup is not straightforward to automate and may need to be done by hand.  If a new collaboration wishes to extend the XML Schemata to mark up data for which no QCDml exists, the process of standardisation of the markup may take a substantial amount of time.

Marking up configurations may be more straightforward, and may be
automated. However, it too involves some amount of post-processing. The checksum needed in the configuration metadata document
is not easy to compute in a parallel program and likewise a unique key;
the configuration LFN; needs to be known in order to create both the
configuration metadata and in order to write a fully ILDG compliant
configuration file as described previously. However, the LFN may not
be known at the time of production. Thus typically configuration
metadata is generated post-production, and the configurations
typically do not contain the LFN on creation. This has to be added on
insertion to the ILDG. 

While much of this activity can be automated, the initial goal of the
computation producing the configuration metadata and the ILDG
compliant configuration at the same time has been sacrificed in order
to agree on other aspects. There is thus scope in the data production
workflow, for data to lay idle for quite some time before being added
to the ILDG with the consequent loss of history and provenance
information. Hopefully future software tools can
alleviate this problem.

Although it was thought that these difficulties will be a major
stumbling block to ILDG participation, in practice metadata creation
proved to be less of a stumbling block than initially expected. The
ensemble metadata typically needs to be created only once, making it
worth the effort and as mentioned previously the workflow for
configuration metadata markup and publication can be substantially
automated. Hence while the in principle issues discussed above remain,
at a practical level the bar for participation in ILDG came not from
the metadata, but rather from maintaining the middleware stack of the
participating organisations such as managing grid security certificate
infrastructure.
%
%

%
%
One aspect of the ILDG to remark upon is that it is most definitely
a volunteer, and altruistic activity. It receives very little in the 
way of funding for itself and is usually piggybacked discretely onto
other grid related projects or to regional grid activities. Correspondingly, 
it can become difficult to maintain effort focused on the ILDG, which 
limits large scale development and essentially forces simple solutions.

%
%
We can contrast the ILDG with some other related work. Other non-ILDG
lattice archives include the Gauge Connection (at NERSC)
\cite{GaugeConnection} and the QCDOC Configuration download site
\cite{QCDOCSite} (at the Brookhaven National Laboratory) which is very
similar in structure to the Gauge Connection and we shall treat the
two identically below. The Gauge Connection was created before the era
of Web Services and Grid services. It hosts files on a single
filesystem and one can download all the configurations over HTTP. The
file format used is an ASCII header followed by a binary data
segment. The header contains rudimentary metadata (e.g. information
about the creators, a checksum, and some derived measurement). Hence
there is no separation between the configuration files and their
metadata like there is in our case. Ensembles are not marked up in
terms of XML at all, but there is some human-readable description for
each one. Authentication and authorisation is done at the Web-Server
level and one needs to register with the site to gain access. This
setup, though very simple has worked very robustly and well. On the other
hand, it becomes harder to search this archive, since there is no
actual metadata catalogue as such. A human must read through a list of
available ensembles until he finds the one he wants from the
description. The Gauge Connection served as a guide to the ILDG
effort. In particular the layout of the data in the binary part of the
Gauge Connection format has been kept in the ILDG data record.

We should also mention in this section the LQCD Archive (LQA)
\cite{LQA} which is maintained at the Center for Computational Sciences
at the University of Tsukuba in Japan. The LQA began development prior
to the ILDG to distribute the data of the CP-PACS collaboration as a
configuration download service similar to the Gauge
Connection. However, upon inception of the ILDG, the LQA was
re-developed to be the front end portal to the data available on the
JLDG.  It currently provides metadata search facilities as well as
HTTP based download which may be useful to users who do not wish to
set up a full grid client infrastructure on their machine. The JLDG
data is of course also available through the usual ILDG client tools
independent of this portal. To use this service, one is required to
register. The portal post a list of publications to which citations
should be made on publication of results that come from the downloaded
datasets.

Download services have proved useful to the community however they
have several shortcomings. They allow downloading primarily through
HTTP which may encounter performance limitations when one considers
downloading entire ensembles, especially since the size of
configurations is expected to increase. There has been no attempt to
provide a common file format. The individual architectures do not lend
themselves to data replication and lack a common security
infrastructure (each requiring separate registrations). That having 
been said, historically the Gauge Connection share their file format
while the LQA as noted above has been extensively redeveloped to complement
rather than contrast with the ILDG.


One can also compare the ILDG to the concept of a Science Gateway.
Quoting from the definition of Science Gateways on the TeraGrid
\cite{TeraGridScienceGateways}, ``A Science Gateway is a community
developed set of tools, applications and data that is integrated via a
portal, or suite of applications, usually in a graphic interface that
is customised to meet the needs of a target community.'' In this sense
the gateways have a broader scope than the ILDG, they can offer codes,
grid services, as well as access to data collections. As an example
we consider the ``Massive Pulsar Surveys Using the Arecibo L-band Feed Array
(ALFA)'' TeraGrid Science Gateway which allows one to brows data
on pulsars and is similar in scope to the ILDG. One can browse pulsar
information, and can download associated data-products. On the other
hand, the SCEC Earthworks Gateway actually allows the
running of earthquake simulations on TeraGrid resources. Both these
gateways can be found at \cite{GatewayList}. 

One unique feature of the ILDG, in contrast to a Science Gateway, is
that the ILDG is the result of a collaboration of collaborations. A
single Science Gateway would typically consist of a single portal
maintained by a group on behalf of a larger community.  This group
then has some freedom (within community limits) in defining internal
formats, markup and can settle on a single set of software tools. The
ILDG instead is a loose federation of existing grids, some of which at
the inception of the ILDG had no grid infrastructure and some of whom
were already heavily invested in their own systems. The worldwide
community had to therefore come together in order to define metadata
standards, middleware operation and thin, easy to implement interfaces
that could then wrap any potentially existing, underlying
infrastructure. Another difference between the ILDG and Science
Gateways may be their philosophy.

\section{Summary and Future Work}
\label{S:conclusions}
%
%
%
In summary, the ILDG is a loosely federated grid-of-grids to facilitate
the sharing of LQCD data worldwide. The technology allows it to
operate across regional grid boundaries, relies on a simple and thin 
layer of middleware standard definitions, and a standardised metadata
markup.

In six years of design and a little over one year of operation, the ILDG effort has brought together the lattice QCD community and has fostered QCD research and collaboration.

Potential future work focuses on several areas including but not limited to data replication, and the storage and mark up of secondary large data such as quark propagators.

%
%

%



\begin{acknowledgements}
  Notice: Authored in part by Jefferson Science Associates, LLC under
  U.S. DOE Contract No. DE-AC05-06OR23177. The U.S. Government retains a
  non-exclusive, paid-up, irrevocable, world-wide license to publish or
  reproduce this manuscript for U.S. Government purposes.
\end{acknowledgements}



\end{document}